\documentclass[structabstract]{aa}  
\usepackage{graphicx}
\usepackage{natbib}
\newcommand{\teff}{\mbox{$T_{\rm eff}$}}
\newcommand{\logg}{\mbox{$\log g$}}
\newcommand{\vsini}{\mbox{$v \sin i$}}
\newcommand{\mictrb}{\mbox{$\xi_{\rm t}$}}
\newcommand{\mactrb}{\mbox{$v_{\rm mac}$}}

\newcommand{\kms}{\mbox{km\,s$^{-1}$}}
\newcommand{\halpha}{\mbox{$H_\alpha$}}
\newcommand{\degree}{\ensuremath{^\circ}}

\def\ms{\hbox{\,m\,s$^{-1}$}}         
\def\m2s2{\hbox{\,m$^{2}$\,s$^{-2}$}} 
\def\kms{\hbox{\,km\,s$^{-1}$}}       
\def\vsini{\hbox{$v$\,sin\,$i$}}      
\def\Msun{\hbox{$M_{\odot}$}}             
\def\Rsun{\hbox{$R_{\odot}$}}
\def\Mjup{\hbox{$\mathrm{M}_{\rm Jup}$}}
\def\Rjup{\hbox{$\mathrm{R}_{\rm Jup}$}}

\def \1s{$1\,\sigma$}

\def \t0{T$_0$}

\usepackage{txfonts}
\begin{document}
\title{WASP-38b: A Transiting Exoplanet in an Eccentric, 6.87d
    Period Orbit}

\author{S. C.  C. Barros \inst{1} \and F. Faedi \inst{1} \and
  A. Collier Cameron \inst{2} \and T. A. Lister \inst{3} \and
  J. McCormac \inst{1} \and D. Pollacco \inst{1} \and E. K. Simpson
  \inst{1} \and B. Smalley \inst{4} \and R. A. Street \inst{3} \and
  I. Todd \inst{1} \and A. H. M. J. Triaud \inst{5} \and I. Boisse
  \inst{6} \and F. Bouchy \inst{6,7} \and G. H\'ebrard \inst{6,7} \and
  C. Moutou \inst{8} \and F. Pepe \inst{6} \and D. Queloz \inst{6}
  \and A. Santerne \inst{8} \and D. Segransan \inst{6} \and S. Udry
  \inst{6} \and J. Bento \inst{9} \and O. W. Butters \inst{10} \and
  B. Enoch \inst{2} \and C. A. Haswell \inst{11} \and C. Hellier
  \inst{4} \and F. P. Keenan \inst{1} \and G. R. M. Miller \inst{2}
  \and V. Moulds \inst{1} \and A. J. Norton \inst{11} \and N. Parley
  \inst{2} \and I. Skillen \inst{12} \and C. A. Watson \inst{1} \and
  R. G. West \inst{10} \and P. J. Wheatley \inst{9} }

\authorrunning{S. C.  C. Barros et al. } \titlerunning{WASP-38b}

\institute{Astrophysics Research Centre, School of Mathematics \&
  Physics, Queen's University Belfast, University Road, Belfast, BT7
  1NN, UK \email{s.barros@qub.ac.uk} \and SUPA, School of Physics \&
  Astronomy , University of St Andrews, North Haugh, St Andrews KY16
  9SS, UK \and Las Cumbres Observatory, 6740 Cortona Drive Suite 102,
  Goleta, CA 93117, USA \and Astrophysics Group, Keele University,
  Staffordshire, ST5 5BG, UK \and Observatoire de Geneve, Universite
  de Geneve, 51 Ch. des Maillettes, 1290 Sauverny, Switzerland \and
  Institut d'Astrophysique de Paris, UMR7095 CNRS, Universite Pierre
  \& Marie Curie, 75014 Paris, France \and Observatoire de
  Haute-Provence, CNRS/OAMP, 04870 Saint-Michel l'Observatoire, France
  \and Laboratoire d'Astrophysique de Marseille, Universit\'e
  d'Aix-Marseille \& CNRS, 38 rue Fr\'ed\'eric Joliot-Curie, 13388
  Marseille cedex 13, France \and Department of Physics, University of
  Warwick, Coventry CV4 7AL, UK \and Department of Physics and
  Astronomy, University of Leicester, Leicester, LE1 7RH \and
  Department of Physics and Astronomy, The Open University, Milton
  Keynes, MK7 6AA, UK \and Isaac Newton Group of Telescopes, Apartado
  de Correos 321, E-38700 Santa Cruz de la Palma, Tenerife, Spain
}

\date{Received August ??, 2010; accepted March ??, 2010}

 
\abstract
{} {We report the discovery of WASP-38b, a long period transiting
  planet in an eccentric $6.871815$ day orbit. The transit epoch is $
  2455335.92050 \pm 0.00074$ (HJD) and the transit duration is $4.663$
  hours.}  {WASP-38b's discovery was enabled due to an upgrade to the SuperWASP-North cameras. We performed a spectral analysis of the host star HD
  146389/BD+10 2980 that yielded $T_{eff} = 6150 \pm 80\,$K, \logg$=4.3
  \pm 0.1$, \vsini=$8.6 \pm 0.4\,$\kms, $M_*=1.16 \pm 0.04\,$\Msun\ and
  $R_* =1.33 \pm 0.03\,$\Rsun, consistent with a dwarf of spectral type
  F8.  Assuming a main-sequence mass-radius relation for the
    star, we fitted simultaneously the radial velocity variations and
  the transit light curves to estimate the orbital and planetary
  parameters.} { The planet has a mass of $2.69 \pm 0.06 $ \Mjup\ and
  a radius of $1.09 \pm 0.03\, $\Rjup\, giving a density, $ \rho_p =
  2.1 \pm 0.1\, \rho_J$.  The high precision of the eccentricity
  $e=0.0314 \pm 0.0044$ is due to the relative transit timing from the
  light curves and the RV shape.  The planet equilibrium temperature
  is estimated at $1292 \pm 33\,$K. WASP-38b is the longest period
  planet found by SuperWASP-North and with a bright host star (V =
  $9.4\,$ mag), is a good candidate for followup atmospheric studies.}

{}


\maketitle

%

\section{Introduction}

Transiting planets are
  important 
because the geometry of these systems gives us a wealth of
information. Photometry during transit allows us to derive the
inclination of the orbit and the radii of both the host star and
planet. Combining this information with radial velocity variations
allows us to derive the absolute mass of the planet and, hence, the
density. Even just an estimation of the bulk density gives us an
insight into the composition of the planet \citep{Guillot2005,
  Fortney2007} and can be used to put constraints on planetary
structure and formation models. These systems also offer a
  potential for measuring planetary emission spectra through
occultation observations (e.g. \citealt{Charbonneau2008}) and we can
gain an insight into the composition of planetary atmospheres using
transit spectroscopy \citep{Charbonneau2002,Vidal-Madjar2003,
  Swain2009}.
 
For these reasons, there are several ground-based surveys searching
for transiting exoplanets, such as HATNet \citep{Bakos2004}, TrES
\citep{alonso2004}, XO \citep{McCullough2005} and WASP
\citep{Pollacco2006}. Currently, there are also two space-based
surveys: CoRoT \citep{Baglin2006} and Kepler \citep{Borucki2010}.
WASP is the most prolific of these surveys having discovered
38 of the 106 known transiting exoplanets. The WASP project consists
of two robotic observatories: one in the Observatorio del Roque de los
Muchachos, La Palma, Canary Islands, Spain and the other in the South
African Astronomical Observatory of Sutherland, South Africa.

In this paper, we report the discovery of WASP-38b, an eccentric giant
planet in a 6.87 day orbit. The candidate was identified in February
2010 in SuperWASP-North data. Radial velocity followup started at the
end of March with {\it FIES} (2.6m NOT). The planetary nature of the
object was established with {\it SOPHIE} (1.93m OHP) and {\it CORALIE}
(1.2m EULER) in May 2010. High precision photometry light curves were
obtained with the Faulkes Telescope North (FTN) and Liverpool
Telescope (LT).

WASP-38b is the 12th longest period of the 106 transiting
  exoplanets reported to date, and the fourth longest period of those
  discovered by ground-based observations. It was discovered after an
  upgrade to the SuperWASP-North cameras which we discuss in section
  2.1. Therefore, WASP-38b is an important object whose properties
add to the known transiting planets parameter space.

\section{Observations}

\subsection{SuperWASP observations}

The SuperWASP-North observatory in La Palma consists of 8 cameras each
with a Canon 200-mm f/1.8 lens coupled to an Andor e2v $2048 \times
2048$ pixel back-illuminated CCD \citep{Pollacco2006}. This
configuration gives a pixel scale of 13.7\arcsec/pixel which
corresponds to a field of view of $7.8\times 7.8$ square degrees per
camera.

In October 2008, we introduced an electronic focus control and
  we also started stabilisation of the temperature of the
SuperWASP-North camera lenses.  Prior to this upgrade, night-time
temperature variations affected the focal length of the
lenses altering the FWHM of stars. This introduced trends in the data
(mimicking partial transits), especially at the beginning and end of
the night when the temperature variation is more extreme. These
effects are not corrected by our detrending algorithms SYSREM
\citep{Tamuz2005} and FTA \citep{Kovacs2005} because they are
position-dependent and do not affect all stars in the same manner. To
reduce this source of systematic noise, heating strips were placed
around each lens so that their temperature is maintained above ambient
at 21 degrees. Besides the stabilisation of the temperature we
  also significantly improved the focus of each of the lenses, which
  now can be done remotely. This upgrade was successful and proved
very important for the discovery of WASP-38b.

\begin{figure}
  \centering
  \includegraphics[width=\columnwidth]{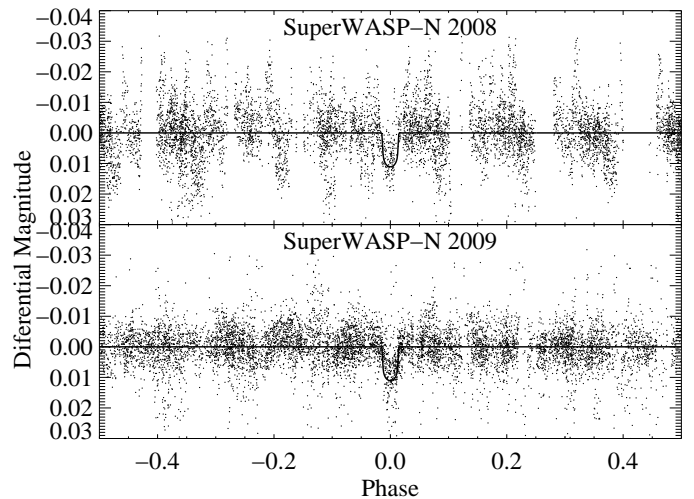}
  \caption{SuperWASP phase folded light curve for WASP-38. On the top
    panel we show the 2008 data and on the bottom we show the 2009
    data after the upgrade.}
  \label{superwasplc}
\end{figure}

The field containing WASP-38 (HD 146389 / BD+10 2980
$\alpha=$16:15:50.36 $\, \delta=$ +10:01:57.3) was observed in the
period between 2008-03-29 and 2008-06-30 by camera 144 (3777 points)
and camera 145 (3278 points).  During this season, our transit search
algorithm \citep{Cameron2006, Cameron2007} did not detect the
transit. In the following year, after the upgrade, observation of this
field continued using the same cameras between 2009-03-30 and
2009-06-30. Camera 144 recorded 5920 observations and camera 145
recorded 2922.  In the 2009 data, transits were detected using both
cameras.  The phase folded light curve using the 2009 SuperWASP data
of WASP-38 is shown in the bottom panel of
Figure~\ref{superwasplc}. We also present the same for the 2008
SuperWASP data in the top panel, showing that the transit is also
visible in the 2008 data.  Comparing both data sets, we conclude that
the somewhat higher rms (8.7mmag) of the 2008 data compared with the
2009 data (6.6mmag) prevented the detection of the transits in the
first observing season.

Hence, we conclude that the upgrade was very successful in reducing
the systematic noise of the SuperWASP-North cameras and allowed the
discovery of a long period transiting planet. WASP-38b is the
longest period transiting exoplanet found by SuperWASP-North. WASP-8b
\citep{queloz2010} has a slightly longer period (P$=8.16\,$days) but
was discovered by WASP South. Ground-based transiting surveys of
exoplanets are biased towards shorter period planets due to their duty
cycle and shorter transits. Reducing the systematic noise will be
important in the discovery of long period and/or smaller radii
planets.

\subsection{Spectroscopic followup}

The first radial velocity measurements of WASP-38 were taken with the
Fibre-Fed Echelle Spectrograph ({\it FIES}) mounted on the 2.56m
Nordic Optical Telescope in La Palma. {\it FIES} was used in medium
resolution mode ($R=46\,000$) with simultaneous ThAr wavelength
calibration. Two observations were made on the nights of 2010-03-29
and 2010-03-30. On 2010 June 08 further nine observations were taken
close to phase zero but out-of-transit.  The observations were reduced
with the FIEStool package and cross-correlated with a high
signal-to-noise spectrum of the Sun to obtain the radial velocities.

The planetary nature of WASP-38b was established with {\it SOPHIE}
mounted on the 1.93m telescope of the Observatoire de Haute Provence
\citep{Perruchot2008,Bouchy2009} and {\it CORALIE} on the 1.2m Swiss
Euler telescope in La Silla \citep{Baranne1996, Queloz2000, Pepe2002}.
Ten measurements were taken by {\it SOPHIE} and 16 by {\it CORALIE}
between 2010 April and July, both achieving a signal-to-noise ratio of
30. The data was reduced with the {\it SOPHIE} and {\it CORALIE}
pipelines, respectively. The radial velocity errors account for the
photon noise plus known systematics in the high efficiency mode
\citep{Boisse2010}.

\begin{figure}
  \centering

  \includegraphics[width=\columnwidth]{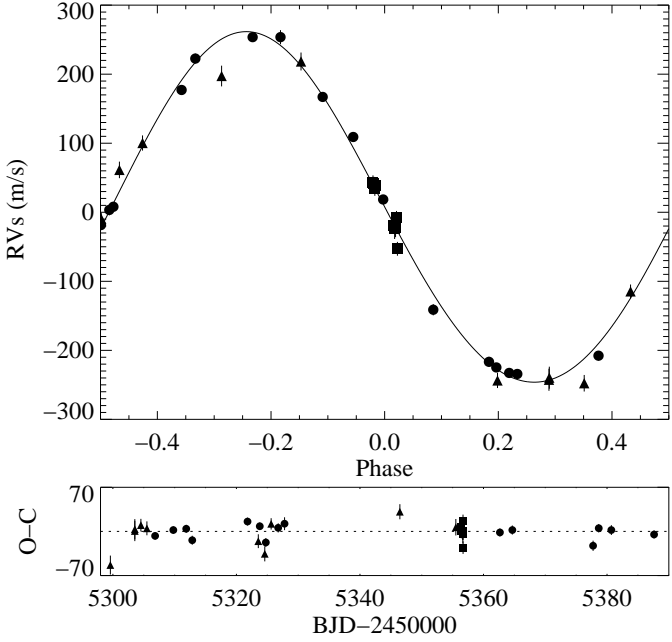}
  \caption{Phase folded radial velocities of WASP-38 obtained with
    {\it FIES} (squares), {\it SOPHIE} (triangles) and {\it CORALIE}
    (circles). The centre-of-mass velocity for each data set was
    subtracted from the RVs. We also show the residuals from the
    orbital fit against time (bottom panel).}
  \label{rvs}
\end{figure}

The radial velocity measurements are given in
Table~\ref{rvobservations}.  In Figure~\ref{rvs}, we show the phase
folded radial velocities from {\it FIES} (squares), {\it SOPHIE}
(triangles) and {\it CORALIE} (circles). We superimpose the best fit
Keplerian model described in section 3.2. In the same figure we show
the residuals from the Keplerian model which show no long term
trend. The semi-amplitude of the radial velocities is $\sim 250 \ms$
consistent with a $2.7\,$\Mjup\ planet in a slightly eccentric orbit.

\begin{figure}
  \centering
  \includegraphics[width=\columnwidth]{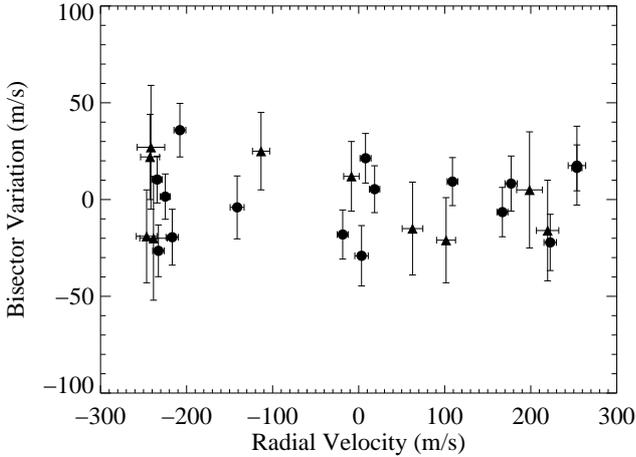}
  \caption{Bisector span measurements for WASP-38 as a function of
    radial velocity for {\it SOPHIE} (triangles) and {\it CORALIE}
    (circles) data.}
  \label{bis}
\end{figure}

A bisector span analysis was performed on the {\it SOPHIE} and {\it
  CORALIE} data and is shown in Figure~\ref{bis}. The bisector span
shows no significant variation nor correlation with the radial
velocities. This suggests that the radial velocity variations are
mainly due to Doppler shifts of the stellar lines rather than stellar
profile variations due to stellar activity or a blended eclipsing
binary.

\begin{table}[ht]
  \centering 
  \caption{Radial velocities of WASP-38}
  \label{table_rv}
  \begin{tabular}{cccc}
    \hline
    \hline
    BJD & RV & $\pm$$1\,\sigma$ & $V_{span}$ \\
    -2\,450\,000 & (km\,s$^{-1}$) & (km\,s$^{-1}$)  & (km\,s$^{-1}$) \\
    \hline
    \multicolumn{3}{c} {{\it FIES}    NOT} \\
    \hline
    5285.6603  &    -9.678   & 0.010&\\
    5286.7164  &    -9.526   & 0.008&\\
    5356.3942  &    -9.800   & 0.010&\\
    5356.4056  &    -9.801   & 0.009&\\
    5356.4170  &    -9.808   & 0.011&\\
    5356.4284  &    -9.804   & 0.010&\\
    5356.6447  &    -9.862   & 0.012&\\
    5356.6561  &    -9.867   & 0.015&\\
    5356.6675  &    -9.865   & 0.014&\\
    5356.6788  &    -9.851   & 0.010&\\
    5356.6902  &    -9.896   & 0.010&\\
    \hline 
    \multicolumn{3}{c}{{\it SOPHIE}     OHP} \\
    \hline
    5299.58942&	-9.510&	0.015&	0.003\\
    5303.54816&	-9.950&	0.016&	0.025\\
    5303.55212&	-9.947&	0.016&	-0.022\\
    5304.53449&	-9.822&	0.010&	0.023\\
    5305.50502&	-9.607&	0.011&	-0.023\\
    5323.54137&	-9.951&	0.011&	0.020\\
    5324.58983&	-9.955&	0.012&	-0.021\\
    5325.62208&	-9.717&	0.009&	0.010\\
    5346.45645&	-9.646&	0.012&	-0.017\\
    5355.52441&	-9.489&	0.013&	-0.018\\
    \hline 
    \multicolumn{3}{c}{{\it CORALIE}     Euler} \\
    \hline
    5306.836479&	-9.5406& 	0.0059&		-0.0225\\ 
    5309.783314&	-10.019& 	0.0059&		-0.0374\\ 
    5311.873800&	-9.8128& 	0.0063&		-0.0570\\ 
    5312.849321&	-9.6171& 	0.0071&		-0.0307\\ 
    5321.796623&	-9.6852& 	0.0062&		-0.0296\\ 
    5323.780775&        -10.0284&	0.0061&		-0.0286\\ 
    5324.762705&	-10.0020&	0.0069&		-0.0031\\ 
    5326.759164&	-9.5716 &	0.0073&		-0.0611\\ 
    5327.789065&	-9.5406 &	0.0102&		-0.0214 \\
    5362.657765&	-9.6272 &	0.0064&		-0.0454\\ 
    5364.669083&	-10.0109&	0.0072&		-0.0584\\ 
    5377.739163&	-9.9354 &	0.0081&		-0.0429\\ 
    5378.657251&	-10.0269&	0.0067&		-0.0654\\ 
    5380.693611&	-9.7909 &	0.0078&		-0.0680\\ 
    5387.614500&	-9.7862 &	0.0064&		-0.0176\\ 
    5404.620483&	-9.7757 &	0.0061&		-0.0336\\ 
    \hline
  \end{tabular}
  \label{rvobservations}
\end{table}

\subsection{Photometric followup}
To better constrain the system parameters, high precision transit
light curves were obtained.  The first photometric followup
observations of WASP-38 were performed on 2010 May 19 using the
LCOGT\footnote{http://lcogt.net} 2.0m FTN located on Haleakala,
Maui. The Spectral instrument was used which contains a Fairchild
$4096\times4096$\, pixel CCD which was binned $2 \times 2$ to give
$0.304\arcsec$ pixels and a field of view of $10\arcmin \times
10\arcmin$. Observations were taken through a Pan-STARRS z filter and
the telescope was defocussed during the observations to prevent
saturation and to increase the exposure time and reduce the effect of
scintillation. The exposure time of the observations was
  $20\,$s. The DAOPHOT photometry package within IRAF was used to
perform object detection and aperture photometry using a 16 pixel
aperture radius. Differential photometry was performed relative to 23
comparison stars within the field-of-view.

Additional photometry was obtained on 2010 June 08 and 15 with a
$18\,$cm Takahashi astrograph in La Palma. The CCD is an Andor
$1024\times1024$ pixel e2v detector with $5.33$\arcsec pixels and
$1.5\degree \times 1.5 \degree$ field of view. The observations were
taken with the $i'$ filter with an exposure time of 15 seconds. Images
were bias and dark subtracted and flat field corrected with standard
IRAF packages. We performed differential photometry relative to 5
comparison stars using DAOPHOT within IRAF.

On the night of 2010 June 15 we also observed WASP-38 with the
high-speed CCD camera RISE mounted on the 2.0m Liverpool Telescope
\citep{rise2008, gibson2008}. RISE has a wideband filter $\sim
  500$ - $700\,$nm which corresponds approximately to V+R. We
obtained 3530 exposures in the $2 \times 2$ binning mode with an
exposure time of 3.7 seconds and effectively no dead time. As usual,
when using RISE for exoplanet transit observations, the telescope was
defocussed by -1.2mm to spread the PSF over a larger number of pixels
thereby increasing the signal-to-noise ratio. This resulted in a FWHM
of $\sim 11$ \arcsec.  The data were reduced using the ULTRACAM
pipeline \citep{Ultracam} which is optimized for time-series
photometry. Each frame was bias subtracted and flat field corrected. We performed differential photometry relative to seven
nearby bright stars, all checked to be non-variable.  We sampled
different aperture radii and chose the aperture radius that minimised
the noise which turned out to be a 24 pixel aperture radius
($13$\arcsec).

\begin{figure}
  \centering
  \includegraphics[width=\columnwidth]{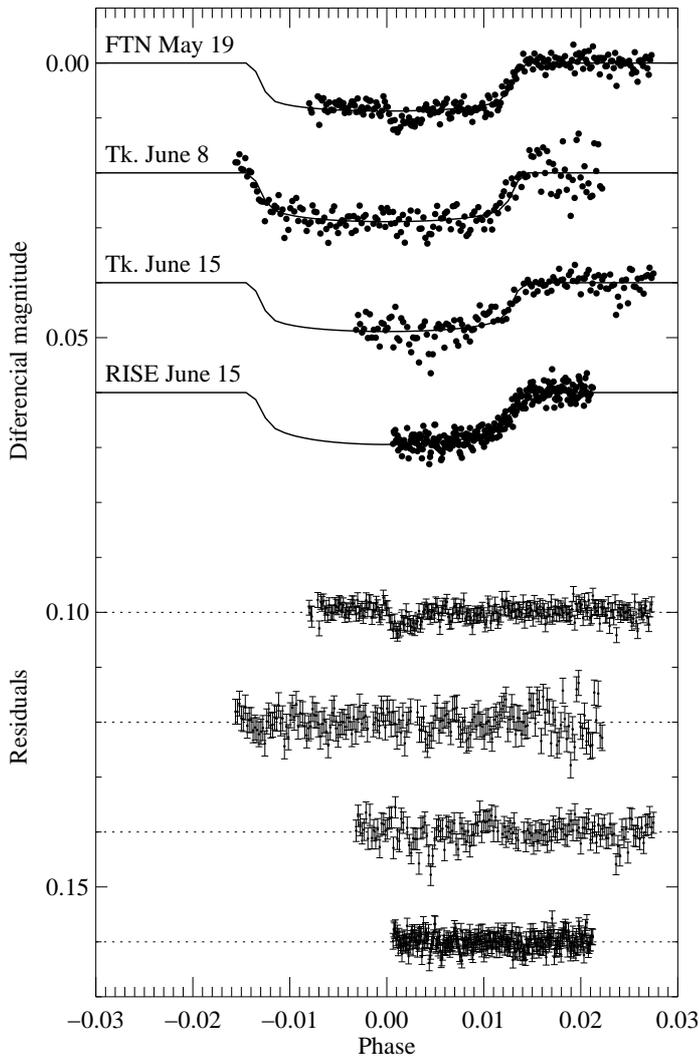}
  \caption{Phase folded light curve for WASP-38. From top to bottom;
    FTN taken on the 2010 May 19, Takahashi (Tk.) astrograph taken on the
    2010 June 8 and 15 and LT/RISE taken on the 2010 June 15. We
    superimpose the best-fit transit model and also show the residuals
    for each light curve on the bottom of the figure. The data were binned and displaced vertically for clarity.}
  \label{photolc}
\end{figure}

The final high precision photometric light curves are shown in
Figure~\ref{photolc} along with the best-fit model described in
section 3.2.

\section{Results and system parameters}

\subsection{Stellar Parameters}

WASP-38 (HD 146389, BD+10 2980) is listed as having spectral type F8
in the HD catalogue \citep{Cannon1921}. This is consistent with that
implied by the value of $B-V = 0.502$ given in the Tycho catalogue
\citep{Hog1997}.

The {\it FIES} spectra were co-added to produce a single spectrum with
a average signal-to-noise of around 200:1. Standard pipeline reduction
products were used in the analysis.

The spectral analysis was performed using the methods given in
\citet{Gillon2009}. The \halpha\ line was used to determine the
effective temperature (\teff), while the Na {\sc i} D and Mg {\sc i} b
lines were employed as surface gravity (\logg) diagnostics. Parameters
obtained from the analysis are listed in
Table~\ref{wasp38-params}. The elemental abundances were determined
from equivalent width measurements of several clean and unblended
lines. A value for microturbulence (\mictrb) was determined from
Fe~{\sc i} using the method of \citet{Magain1984}. The quoted error
estimates include that given by the uncertainties in \teff, \logg\ and
\mictrb, as well as the scatter due to measurement and atomic data
uncertainties.

The projected stellar rotation velocity (\vsini) was determined by
fitting the profiles of several unblended Fe~{\sc i} lines. A value
for macroturbulence (\mactrb) of 4.9 $\pm$ 0.3 \kms\ was assumed,
based on the tabulation by \citet{Gray2008}, and an instrumental FWHM
of 0.13 $\pm$ 0.01 \AA\ was determined from the telluric lines around
6300\AA. A best-fit value of \vsini\ = 8.6 $\pm$ 0.4~\kms\ was
obtained.

We estimated the distance by comparing the V magnitude (V = 9.447)
taken from Tycho \citep{Hog1997} with the absolute magnitude of a
F8-type star from \citet{Gray1992}.

\begin{table}[ht]
  \centering 
  \caption{Stellar parameters of WASP-38 from spectroscopic analysis.}
  \begin{tabular}{cc}
    RA(J200)     & 16:15:50.36 \\
    DEC(J2000)   & +10:01:57.3 \\
    V(mag)       & 9.447  $\pm$ 0.024 \\
    \teff        & 6150 $\pm$ 80~K \\
    \logg\ [cgs] & 4.3 $\pm$ 0.1 \\
    \mictrb      & 1.4 $\pm$ 0.1 \kms \\
    \vsini       & 8.6 $\pm$ 0.4 \kms \\
    log A(Li)    &   1.93 $\pm$ 0.08 \\
    Mass [\Msun]  &   1.16 $\pm$ 0.09 \\
    Radius [\Rsun]&   1.26 $\pm$ 0.17 \\
    Spectral Type &   F8 \\
    Distance      &    110 $\pm$ 20~pc \\
    &  \\
    {[Fe/H]}   &$-$0.12 $\pm$ 0.07 \\
    {[Na/H]}   &$-$0.07 $\pm$ 0.07 \\
    {[Mg/H]}   &$-$0.03 $\pm$ 0.07 \\
    {[Si/H]}   &$-$0.01 $\pm$ 0.04 \\
    {[Ca/H]}   &$+$0.00 $\pm$ 0.13 \\
    {[Sc/H]}   &$-$0.03 $\pm$ 0.16 \\
    {[Ti/H]}   &$-$0.06 $\pm$ 0.12 \\
    {[V/H]}    &$-$0.17 $\pm$ 0.09 \\
    {[Cr/H]}   &$-$0.08 $\pm$ 0.11 \\
    {[Mn/H]}   &$-$0.22 $\pm$ 0.12 \\
    {[Co/H]}   &$-$0.17 $\pm$ 0.21 \\
    {[Ni/H]}   &$-$0.14 $\pm$ 0.07 \\
    \\
  \end{tabular}
  \label{wasp38-params}
  \newline {\bf Note:} Mass and radius estimate using the
  \citet{Torres2010} calibration. Spectral type from HD Catalogue.
\end{table}

\subsection{Planet parameters}

To determine the planetary and orbital parameters, we fitted all the
photometry and radial velocity measurements simultaneously.  Our model
is an updated version of the Markov-Chain Monte Carlo (MCMC) fitting
procedure described by \citet{Cameron2007} and \citet{Pollacco2008}.
Our global fit uses the \citet{Mandel2002} transit model parametrised
by the transit epoch $T_0$, orbital period $P$, impact parameter $b$,
transit duration $T_T$ and squared ratio of planet radius to star
radius $ (R_p/R_*)^2 $. For each photometric data set, we include the
non-linear limb darkening coefficients for the respective filter based
on the tables of \citet{Claret2000, Claret2004}. The Keplerian model
for the host star's reflex motion is parametrised by the
centre-of-mass velocity $\gamma$, the radial velocity amplitude $K$,
the orbital eccentricity $e$ and the longitude of the periastron $w$.

The main difference in the new version of our MCMC code is that the
stellar mass is no longer an input parameter and is estimated from
\teff, $\rho_*$ and {[Fe/H]} using the calibration of
\citet{Torres2010} as described in \citet{Enoch2010}. While \teff\,
and {[Fe/H]} are input parameters derived from spectral fitting
(Table~\ref{wasp38-params}), $\rho_*$ is estimated at each point in
the chain directly from the light curves.

Due to the poor quality of the only complete transit of WASP-38b we
imposed a main-sequence mass-radius relation for the parent star,
i.e. $R_*=M_* ^{0.8}$ \citep{Seager2003,Cox2000} in our
global fit.  To better constrain the system parameters, a high
precision complete transit light curve is needed.  Unfortunately, due
to its long transit duration there are not many full transits
observable for this target and the only full transit visible from La
Palma this season failed due to technical issues.

The system parameters of WASP-38 and the 1$\sigma$ uncertainties
derived from the MCMC analysis are given in Table~\ref{mcmc}.
WASP-38b is a $ 2.691\,$\Mjup\, giant planet with an eccentric
($e=0.031$) $6.87$ day orbit. The planet radius is $1.09$ \Rjup, and
hence, it has a high density of $2.06\, \rho_J $.

\begin{table}[h]
  \centering 
  \caption{WASP-38 system parameters.}
  \label{mcmc}
  \begin{tabular}{lccl}
    \hline
    \hline
    Parameter & Value  \\
    \hline
    Transit epoch $T_0$ [HJD] & $ 2455335.9205  \pm 0.00074 $ \\
    Orbital period $P$ [days] & $  6.871815  ^{+ 0.000045}_{- 0.000042} $ \\
    Planet/star area ratio $ (R_p/R_*)^2 $ & $ 0.00712 \pm 0.00018  $  \\
    Transit duration $T_T$ [days] & $ 0.1942 ^{+ 0.0018}_{- 0.0019} $ \\
    Impact parameter $ b $ [$R_*$]  & $ 0.066^{+ 0.093}_{-0.046} $  \\
    Orbital inclination $ I $ [degrees] & $ 89.69^{+0.30}_{- 0.25} $ \\
    &    &      &  \\
    Stellar reflex velocity $K$ [\ms] & $ 253.9 \pm 2.4 $ \\
    Orbital semimajor axis $ a $ [AU] & $ 0.07522 ^{+ 0.00074}_{- 0.00075} $ \\
    Orbital eccentricity  $ e $ & $ 0.0314 ^{+ 0.0046}_{- 0.0041}  $  \\
    Longitude of periastron  $ \omega $[degrees] & $- 16.^{+ 18}_{-17 } $  \\
    &    &      &  \\
    Stellar mass $ M_* $ [\Msun] & $ 1.203 \pm 0.036  $  \\
    Stellar radius $ R_* $ [\Rsun]  & $ 1.331 ^{+ 0.030}_{- 0.025}  $  \\
    Stellar surface gravity $ \log g_* $ [cgs]   & $ 4.250 ^{+ 0.012}_{- 0.013} $  \\
    Stellar density $ \rho_* $ [$\rho_\odot$] & $ 0.509 \pm 0.023  $ \\
    &    &      &  \\
    Planet mass  $ M_p $ [\Mjup]  & $ 2.691  \pm  0.058 $  \\
    Planet radius $ R_p $ [\Rjup]  & $ 1.094 ^{+ 0.029}_{- 0.028} $  \\
    Planet density  $ \rho_p $ [$\rho_J$]  & $ 2.06 \pm 0.14 $  \\
    \hline
  \end{tabular}
 
\end{table}

\subsection{Eccentricity}

The current version of our MCMC code uses the parameters
$\sqrt{e}\cos\omega$ and $\sqrt{e}\sin\omega$ as jump parameters. This
scaling allows the parameter space to be explored efficiently at small
eccentricities, as recommended by \citet{Ford2006}, but ensures a
uniform prior on $e$ (Collier Cameron et al 2010, in prep).

From our global MCMC fit we derived an eccentricity of $ 0.0314 ^{+
  0.0046}_{- 0.0041} $ which although being very small is significant
at 7 $\sigma$.  Given the small eccentricity we also tried fitting a
circular orbit for WASP-38. The $\chi^2$ value for the eccentric model
fit is 77 while the $\chi^2$ value for the circular model is 143. The
eccentric model is parametrised by six parameters: $\gamma$, $K$, $e
\cos \omega$, $e \sin \omega$ and two offsets to account for the shift
between the zero points of the {\it FIES}, {\it SOPHIE} and {\it
  CORALIE}. The first two {\it FIES} points were excluded from the fit
due to contamination from the moon hence we used a total of 35
RVs. Therefore, the Lucy and Sweeney test \citep{Lucy1971} give a
99.99\% probability for the eccentric orbit.

Interestingly if we fit only the radial velocities, the eccentricity
is not significantly detected and the solution is compatible with a
circular orbit. A more careful analysis of our data revealed that our
high sensitivity to the eccentricity comes from the timing of the
transit relative to the RV curve which places a tight constraint on $e
\cos \omega = 0.0293\pm 0.0036 $ while $e \sin \omega$ is consistent
with zero. This is contrary to the common assumption that the
eccentricity is almost solely constrained by the RV curve. The transit
of WASP-38 occurs $\sim 1.7$ hours earlier than what was expected from
the RVs if the orbit was circular. The timing shift is consistent for
all of the followup light curves.

\section{Discussion}
The newly discovered planet WASP-38b is quite similar to the other
currently known long-period transiting planets. It is massive ($
2.69\,$\Mjup), has an eccentric orbit ($e=0.031$) and does not suffer
from the radius anomaly. For an updated list of the properties of
these objects see \citet{Kovacs2010}. Only eleven out of the 106
transiting exoplanets have orbital periods longer than WASP-38b. Of
these, five have been discovered by the CoRoT mission, two were found
by Kepler \citep{Holman2010}, two were found in radial velocity
surveys (HD17156b \citep{Fischer2007,Barbieri2007} and HD80606b
\citep{Naef2001, Moutou2009,Fossey2009,Garcia-melendo2009}) and the
remaining three are WASP-8b, HAT-P-15b \citep{Kovacs2010} and
HAT-P-17b \citep{Howard2010}. Therefore, WASP-38b is the forth
exoplanet with a period longer than six days discovered in a
ground-based transit survey.

The low number of transiting planets with periods longer than five
days is mostly due to selection effects. It is widely known that the
transit probability decreases with period. Moreover, for ground-based
surveys (which are responsible for the discovery of 72\% of the
transiting planets), the detection probability also steeply decreases
with period. This is due to the longer duty cycle of the transits and
longer transit duration coupled with the restricted observing time
from a single site on Earth. To increase the duty cycle of the
observations, telescope networks spread in geographic latitude or
space-based surveys are needed. In the case of WASP-38b it was very
important to reduce the systematic noise which ultimately allowed the
discovery of the planet.  However, the selection effects might be
hiding a real decrease of the number of planets at longer periods. In
fact, from radial velocity surveys there appears to be a depletion of
planets between $ \sim 0.1-1$AU \citep{Udry2003}.

The low lithium abundance, $log A(Li) = 1.93$ points to an age of $>
5$ Gyr for WASP-38 \citep{Sestito2005}. However, it has been shown
\citep{Israelian2009} that stars with planets have an under-abundance
of lithium compared with stars without planets. Therefore, in this
case, the lithium abundance might be overestimating the age. In fact,
if we estimate the age from the rotation period ($ \sim 7.5$ days), we
obtain $ \sim 1$ Gyr from \citet{Barnes2007} (using Tycho
B-V=0.5). Unfortunately, our light curves are not good enough to
constrain the radius of the star.  As mentioned above, we had to
assume the mass-radius relation for the main-sequence in our
parameters fit. Hence, we cannot use the stellar radius to calculate
the isochrone age. Further observations are needed in order to better
constrain the age and the evolutionary status of the star.

WASP-38b is very dense but not atypically so for a massive planet.
Its equilibrium temperature is $1292 \pm 33\,$K which is quite hot for
a long period planet due to its ``hot'' F8 host star that has a
luminosity $\sim 2.4 L_{\odot}$. To receive the same flux in our solar
system the planet would have to be at $0.049$AU from the Sun.
WASP-38b is a ``pL'' class planet according to the classification of
\citet{Fortney2008}. 
Therefore, we expect an efficient re-distribution of heat from the day
side to the night side of the planet and no temperature inversion in
the atmosphere.

A better insight on the planet composition will require a better
constraint on its age.  According to the \citet{Fortney2007} models,
if WASP-38b is $ \sim 1$Gyr old it might have a substantial core with
a mass up to 100 Earth masses. However, if the planet is much older
($4.5$Gyr) its radius is consistent with a hydrogen/helium coreless
planet. A better estimation of the radius of the planet will also help
constrain its composition. Given that the star is metal poor, it would
be interesting to determine the existence of a core.

The Safronov number for WASP-38b is $ \sim 0.3$
\citep{Hansen2007}. With the exception of CoRoT-4b, all the transiting
planets with a period longer than WASP-38b have Safronov numbers
larger than $0.3$ and hence they do not belong to either of the
classes proposed by \citet{Hansen2007}.

Due to its long period it is not surprising that WASP-38b is slightly
eccentric. As discussed above, the eccentricity however small is
significant.  The circularization timescale is given by
\citep{Goldreich1966,Bodenheimer2001}:
\begin{equation}
  \tau_{\mathrm{CIR}} \approx 0.63 \left(\frac{Q_p}{10^6}\right) \left(\frac{M_p}{\Mjup}\right)  \left(\frac{\Msun}{M_*}  \right) ^{3/2}  \left(\frac{a}{10\Rsun}  \right)^{13/2 } \left(\frac{\Rjup}{R_p}  \right) ^{5}\, Gyr.
\end{equation}
From studies of binary stellar evolution \citep{Meibom2005} and from
our Solar System \citep{Goldreich1966,Peale1999}, the tidal
dissipation parameter is $Q_p=10^5-10^6$. Therefore, for WASP-38b the
circularisation timescale is between $\sim 1.9-19\,$Gyr which is
consistent with our constraint on the age of WASP-38 and can be
compared to the main-sequence lifetime for an F8 star ($\sim
9\,$Gyr). Therefore, depending on the value of $Q_p$, WASP-38b's orbit
might never circularise, as it appears to be at the limit of
circularisation. All longer period transiting planets are eccentric or
the eccentricity has been fixed to zero. A better constraint on the
age of WASP-38 might be important to constraint the tidal dissipation
parameters, so further studies are encouraged.

WASP-38 is a bright star (V= 9.4 mag) and therefore is a good
candidate for followup observations. The secondary transit is
predicted to be at phase $0.520 \pm 0.002$, $T_0 = 2455456.3055 \pm
0.015$ and have a duration of $ 274.5 \pm 5.0$ minutes. Next year, the
only full transit visible from La Palma is on 2011 April 21.
Observations of a spectroscopic transit to measure the
Rossiter-McLaughlin effect were obtained in 2010 June by one of our
co-authors and will be presented elsewhere (Simpson et al. 2010, in
prep.).

\begin{acknowledgements}
  The WASP Consortium consists of astronomers primarily from Queen's
  University Belfast, St Andrews, Keele, Leicester, The Open
  University, Isaac Newton Group La Palma and Instituto de Astrofsica
  de Canarias. The SuperWASP-N camera is hosted by the Issac Newton
  Group on La Palma. We are grateful for their support and
  assistance. Funding for WASP comes from consortium universities and
  from the UK’s Science and Technology Facilities Council. Based on
  observations made with the Nordic Optical Telescope, operated on the
  island of La Palma jointly by Denmark, Finland, Iceland, Norway, and
  Sweden, in the Spanish Observatorio del Roque de los Muchachos of
  the Instituto de Astrofisica de Canarias.  Based on observations
  made at Observatoire de Haute Provence (CNRS), France and at the ESO
  La Silla Observatory (Chile) with the {\it CORALIE} Echelle
  spectrograph mounted on the Swiss telescope. The research leading to
  these results has received funding from the European Community's
  Seventh Framework Programme (FP7/2007-2013) under grant agreement
  number RG226604 (OPTICON). FPK is grateful to AWE Aldermaston for
  the award of a William Penny Fellowship. The RISE instrument mounted
  in the Liverpool Telescope was designed and built with resources
  made available from Queen’s University Belfast, Liverpool John
  Moores University and the University of Manchester.  The Liverpool
  Telescope is operated on the island of La Palma by Liverpool John
  Moores University in the Spanish Observatorio del Roque de los
  Muchachos of the Instituto de Astrofisica de Canarias with financial
  support from the UK Science and Technology Facilities Council.  We
  thank Tom Marsh for the use of the ULTRACAM pipeline. SCCB is
  grateful to Catherine Walsh for proofreading this paper.
\end{acknowledgements}

\bibliographystyle{bibtex/aa} \bibliography{susana}

\end{document}